\pdfoutput=1
\documentclass[epj]{svjour}
\usepackage{cite,amsmath,amssymb}
\usepackage{multirow,graphicx,pstricks,upgreek}
\usepackage[title]{appendix}
\usepackage{cuted}
\newcommand{\dd}{\mbox{d}}
\renewcommand{\vec}[1]{\mathbf{#1}}

\newcommand{\EE}{{\cal E}}
\newcommand{\FF}{{\cal F}}

\newcommand{\xiv}{\mbox{\boldmath$\xi$}}
\begin{document}
\title{Dynamic structure factor of a stiff polymer in a glassy solution}

\author{J. Glaser\inst{1} \and O. Hallatschek \inst{2} \and K. Kroy\inst{1}$^,$\inst{3}
}                     
%
%
\institute{Institut f\"ur Theoretische Physik, Universit\"at Leipzig,
  PF 100920, D-04009 Leipzig \and Lyman Laboratory, Harvard
  University, Cambridge MA 02138, USA \and Hahn-Meitner Institut
  Berlin, Glienicker Str. 100, D-14109 Berlin}
\date{Received: date / Revised version: date}
%
\abstract{We provide a comprehensive overview of the current
  theoretical understanding of the dynamic structure factor of stiff
  polymers in semidilute solution based on the wormlike chain (WLC)
  model. We extend previous work by computing exact numerical
  coefficients and an expression for the dynamic mean square displacement (MSD)
  of a free polymer and compare various common approximations for the hydrodynamic
  interactions, which need to be treated accurately if one wants to extract
  quantitative estimates for model parameters from experimental data. A
  recent controversy about the initial slope of the dynamic structure
  factor is thereby resolved. To account for the interactions of the
  polymer with a surrounding (sticky) polymer solution, we analyze
  an extension of the WLC model, the glassy wormlike chain (GWLC), which
  predicts near power-law and logarithmic long-time tails in the dynamic
  structure factor.
} 
\maketitle

\section{Introduction}
Semiflexible polymers constitute a wide class of biologically and
technologically important macromolecules. In contrast to flexible
polymers, their persistence length $\ell_p$, which is the length scale
defining the cross-over from a locally rodlike to a globally coiled
conformation, defines a mesoscale much larger than the monomer size.
Recently, reviving interest in applying scattering techniques to
semiflexible polymers results from the need for reliable methods for
analyzing the mechanical properties of biopolymers such as actin
\cite{LeGoff2002,Semmrich2007}, microtubules
\cite{Pampaloni2006}, intermediate filaments \cite{Hohenadl1999}, DNA
\cite{Winkler2006}, fibrin \cite{Arcovito1997,Pierno2006} or plant
cell wall polysaccharides \cite{Vincent2007}, but also peptide fibrils
\cite{Carrick2005}, wormlike micelles \cite{Buchanan2005}, and
others. The mechanical properties of biopolymers in solution have also
been studied extensively by a number of real space methods such as
particle tracking, single molecule techniques and microrheology. These
provide a more intuitive access to the studied molecules than the
classical scattering methods: even without knowing exactly what to
expect, one obtains telling pictures and videos of the system of
interest. Scattering methods, on the other hand, have not only the
advantage of being applicable also to molecules that are too small to
be visualized under the microscope, they moreover allow for a very
efficient quantitative determination of model parameters, with the
otherwise laborious ensemble averaging automatically included.

Clearly though, their useful application requires a faithful and
accurate representation of the analyzed system within a theoretical
model that serves to analyze the data.  The standard model for the
theoretical analysis of semiflexible polymers is the wormlike chain
(WLC) model, which represents the polymer as an inextensible space
curve with a bending rigidity $\kappa = k_BT \ell_p$ (in three space
dimensions). Although this model is generally quite complicated to
solve, essentially all linear and non-linear dynamical properties of
fundamental interest -- in and out of equilibrium -- can be obtained
via a systematic perturbation calculation around the weakly bending
rod limit \cite{Hallatschek2007}, leaving (in our opinion) little
incentive for considering models involving uncontrolled
approximations. All our computations, which will be detailed below,
are based on the weakly bending limit and all of our results are
derived in (and require for their quantitative validity) the limit of
large $q\ell_p\gg 1$ and $qL\gg1$.

While dynamic light scattering (DLS) has proven a versatile tool for
the study of semiflexible polymer in the past
\cite{Fujime1985,Farge1993,Janmey1994,Gotter1996,Hohenadl1999}, we
feel that the surge of applications and the improving experimental
accuracy call for a more precise evaluation of the theoretical
predictions than previously accomplished. Though dynamical scaling
regimes of isolated semiflexible polymers have been known for a while
\cite{Kroy2000}, a thorough quantitative analysis is presently still
limited by (1) approximations involved in dealing with the
hydrodynamic interactions; and (2) serious deviations of typical
measured structure factors from the idealized theoretical predictions
for an isolated single polymer. The latter are usually due to direct
polymer-polymer interactions, which available models fail to predict
quantitatively. The present contribution therefore extends previous
theoretical work for the dynamic structure factor of stiff polymers in
solution by various authors
\cite{Farge1993,Kroy1997,Kroy2000,Liverpool2001,Nyrkova2007} in
several directions. First, we compute via diverse routes numerical
coefficients that depend on the explicit representation of the solvent
hydrodynamics and are needed for extracting reliable values for the
polymer backbone diameter and the persistence length from the
scattering curves. A new expression for the dynamic mean square displacement
(MSD) of a free chain fully including the effect of hydrodynamic interactions is given.
The analysis sheds some light on the quantitative
reliability of the method and settles a recent controversy concerning
the initial slope of the dynamic structure factor
\cite{Kroy1997,Liverpool2001,Nyrkova2007} in favor of
refs.~\cite{Kroy1997,Nyrkova2007}. Further, we extend the analysis of
the dynamic scattering functions to the practically important case
that the interaction of the scattering polymer with the surrounding
polymer solution is not negligible. While previous discussions of
steric cage and entanglement effects were generally based on the tube
model \cite{Janmey1994,Kroy1997,Morse2001}, we introduce a new,
superior model, called the glassy wormlike chain (GWLC)
\cite{Kroy2007}.  Very recently, thanks to experimental
progress, pronounced logarithmic tails due to collective network
dynamics have been discovered in actin solutions at low temperature
\cite{Semmrich2007} (see also Fig.~\ref{sqtfit}).
They cannot be explained within the common
tube model and have been attributed to a glassy slowdown of the single
polymer relaxation that, according to the GWLC, results from an
exponential stretching of the long-time relaxation spectrum of an
`ordinary' WLC. The GWLC can potentially accommodate steric (free
volume) as well as sticky (enthalpic) \cite{Kroy2007} interactions between the
polymers.  As we show below, it can account for the observed
logarithmic tails of the dynamic structure factor.
It turns out that the presence
of the surrounding polymer matrix not only affects the long-time tails
of the structure factor but causes severe stretching already at
intermediate times. It is therefore indispensable to include the newly
derived theoretical expressions in any attempt to extract a reliable
value for the persistence length from scattering data whenever the
tails of the structure factor exhibit any noticeable deviations from
the ideal stretched exponential form predicted for a single free
polymer in solution. In practice, this remark concerns essentially all
but the most dilute biopolymer solutions.

The remainder of the paper is organized as follows. For convenience,
the second section contains a brief overview of our main results,
which are compared to previous predictions. The third section
summarizes the theoretical model, both the WLC and GWLC model, as far
as needed for the calculation of the dynamic structure factor, detailed
in section~\ref{resultslong}.

\section{Summary of our results}
The coherent dynamic structure factor $S(q,t)$ of a single polymer
chain is defined as
\begin{equation}
S(q,t)=\frac{1}{Ld} \int\dd s\, \dd s'\, \langle e^{i \vec q (\vec r(s',t)-
\vec r(s,0))} \rangle,
\label{cohsf}
\end{equation}
where $L$ is the polymer length, $d$ is the size of a monomer, or,
more accurately, the backbone diameter, and $\vec r(s,t)$ are the monomer positions.
The brackets $\langle \dots \rangle$ denote an ensemble average.
Experimentally, the dynamic structure
factor is obtained from the intensity time-autocorrelation function
measured in dynamic scattering experiments such as DLS with the help of the Siegert
relation, or by neutron spin-echo spectroscopy.

\begin{figure}
\resizebox{\columnwidth}{!}{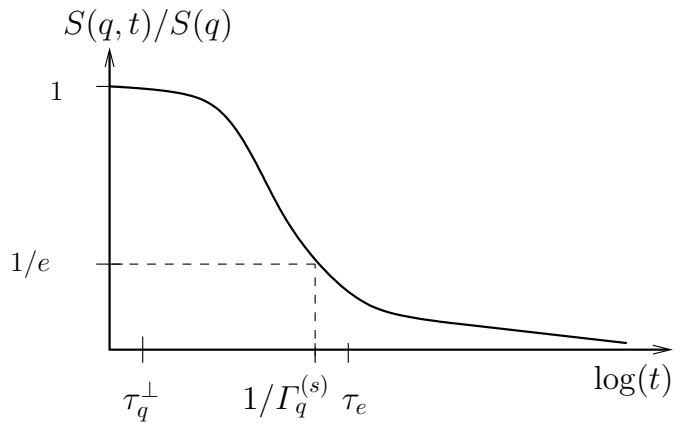}
\caption{General shape of the dynamic structure factor of a single
  semiflexible polymer in solution. The characteristic time
  $\tau_q^\perp=(\zeta_\perp/\kappa) q^{-4}$ separates the initial
  decay regime with decay rate $\Gamma_q^{(0)}$ from the stretched
  exponential tail with decay rate $\Gamma_q^{(s)}$, which reflects
  the structural relaxation. For times $t\gg \tau_e$ longer than the
  entanglement time $\tau_e\equiv (\zeta_\perp/\kappa) L_e^4/\pi^4$,
  the decay may exhibit a more extreme (logarithmic) stretching due to
  the slowdown of the relaxation of wavelengths longer than $L_e$ by
  their interactions with the surrounding (sticky) polymer
  matrix.}\label{fig:schematic}
\end{figure}

One has to distinguish three intermediate regimes of the structure factor
of a single polymer in solution, as schematically indicated in
fig.~\ref{fig:schematic}. The characteristic time scale
$\tau_q^\perp=(\zeta_\perp/\kappa) q^{-4}$ ($\zeta_\perp$ is the
transverse friction per unit length and $q$ the length of the
scattering vector), which was already identified in
refs.~\cite{Kroy1997,Kroy2000,Liverpool2001,Nyrkova2007}, separates
the initial decay regime from the stretched exponential regime. While
the former is due to the rapid random wiggling of the contour dictated
by the thermal forces and only weakly (via the hydrodynamic
interactions) dependent on the mechanical properties of the polymer,
the latter bears the signature of the systematic structural relaxation
driven and controlled by the bending rigidity. Also, while during the
initial decay regime the spatial interference of scattered light from
distant monomers along the contour contributes, there is no
distinction between coherent and incoherent structure factor at late
times \cite{Kroy2000}. Ultimately, the structure factor depends on time
merely implicitly via the dynamic transverse mean-square displacement (MSD)
\begin{equation}
  \delta r_\perp^2(t) \equiv \overline{\langle [\vec r_\perp(s,t)-\vec r_\perp(s,0)]^2
  \rangle}
\end{equation}
of an average bulk monomer (the over-bar denotes a spatial average),
which we assume to be Gaussian distributed, justified by the equation of
motion of a single polymer:
\begin{equation}
  S(q,t) \sim S(q,0) \exp[-q^2 \delta r_\perp^2(t)/4] \qquad (t \gg \tau_q^\perp).
\label{sqtmsd}
\end{equation}
Which (and how much) of these different regimes is detected in a
particular experiment depends sensitively on $q\ell_p$. For large
$q\ell_p\gg 1$ the stretched exponential tail dominates, while it
vanishes for $q\ell_p\approx 1$ \cite{Hohenadl1999}.
Finally, under practical circumstances, polymer
solutions used for scattering
experiments are rarely dilute but rather semidilute to achieve a
decent scattering intensity
Hence, even if the scattering wavelength can be
made smaller than the mesh size, interactions
between the polymers usually become noticeable for times longer than
the entanglement time $\tau_e$, also indicated in fig.~\ref{fig:schematic}.

\paragraph{Short times ($t \ll \tau_q^\perp$)}  
For $\tau \ll \tau_q^\perp$ the dynamic structure factor approaches a
simple exponential, $S(q,t)/S(q,0)$ $\sim \exp(-\Gamma_q^{(0)} t)$.
Below, we present a new discussion of this so-called initial decay
regime, which yields new results.  Our derivation of the initial decay
rate
\begin{equation}
  \Gamma_q^{(0)} = \frac{k_B T q^3}{6 \pi^2 \eta} [C - \log(q d)]\qquad
  (\tau_q^\| \ll t \ll \tau_q^\perp) 
\label{inidec}
\end{equation}
qualitatively confirms some previous results
\cite{Kroy1997,Nyrkova2007}, while ruling out others
\cite{Liverpool2001}. We also compute numerical values for the
constant $C$ explicitly via diverse routes and using various slightly
different forms of hydrodynamic mobility tensor, which are summarized
in table \ref{cpartab}. While the value in the lower right corner
should be considered the most accurate result, the spread of the
numerical values serves well to demonstrate the sensitivity of the
initial decay rate to the details of the hydrodynamic modelling, thus
suggesting some reservation against a too naive identification of the
parameters $d$ extracted from fits to experimental data with the
physical backbone diameter.

\begin{table*}
  \centering\begin{tabular}{|l||c|c|c|c|}
    \hline
    Hydrodynamic interactions&\multirow{2}{1cm}{Oseen}&\multirow{2}{1.5cm}{transverse mobility}
    &\multirow{2}{2.5cm}{Oseen + diagonal terms} 
    &\multirow{2}{2.5cm}{RP  + diagonal terms}\\
    \cline{1-1}
    equations of motion &&&&\\
    \hline
    \hline
    Smoluchowski&5/6 \cite{Kroy1997}&$4/3-\gamma_E$&$8/3-\gamma_E$&-\\
    \hline
    Langevin&\multicolumn{2}{c|}{$4/3 - \gamma_E$}
    &$8/3-\gamma_E$&$17/12+4/3-\gamma_E$\\
    \hline
\end{tabular}
\vspace{.5cm}
\caption{Values of the hydrodynamic constant $C$ appearing in the
  initial decay rate eq.~\eqref{inidecshort} for different methods
  of calculation and different mobility tensors: original Oseen
  tensor, Oseen tensor with longitudinal degrees of freedom projected
  out, Oseen tensor with diagonal (Stokes) friction term and Rotne-Prager
  (RP) mobility function with Stokes friction term.}
\label{cpartab}
\end{table*}

Taking into account longitudinal contour fluctuations, the authors of
\cite{Liverpool2001,Nyrkova2007} introduced a new time scale
$\tau_q^\| \simeq (q \ell_p)^{-4} \tau_q^\perp$. For times
$t\ll\tau_q^\|$ the initial slope is predicted to increase by a factor
of two due to contributions from longitudinal modes, which are
sub-dominant for any fixed time $t$ in the limit $q\ell_p\to \infty$
but become important for any fixed value of $q\ell_p$ upon taking the
limit $t\to0$. Instead of repeating the extended discussion, we refer
the reader to \cite{Nyrkova2007,Hallatschek2007} for the longitudinal
dynamics.  For molecules that are reasonably stiff on the length scale
$q^{-1}$ (which is a requirement for the application of the weakly
bending rod approximation on which our whole discussion is based),
only the contribution due to transverse motions is measurable in
practice because of the double scale separation
\begin{equation}
  \Gamma_q^{(0)}\tau_q^\| =  (q \ell_p)^{-4} \Gamma_q^{(0)} \tau_q^\perp
  \ll  \Gamma_q^{(0)} \tau_q^\perp 
  \simeq (q \ell_p)^{-1} \ll 1 \;.
\end{equation}
The presence of longitudinal contour fluctuations has to be considered
in the limit $q\ell_p\to 1$, though.

\paragraph{Long times ($t \gg \tau_q^\perp$)}
For long delay times the dynamic structure factor is given by
eq.~\eqref{sqtmsd} \cite{Kroy1997}. For the WLC, one calculates a
time-dependent transverse `correlation length' $\ell_\perp(t)$ $=(\kappa
t/\zeta_\perp)^{1/4}$ that characterizes the wavelength of the modes
equilibrating within the time $t$. The transverse friction coefficient per length
\begin{equation}
\zeta_\perp\simeq \frac{4 \pi \eta}{-\log[d/\overline\ell_\perp(t)]},
\end{equation}
owes its time dependence to the hydrodynamic interactions that
facilitate cooperative relaxations over an effective polymer length
$\overline\ell_\perp(t)\approx \ell_\perp(t)$. [For a more precise definition the
reader is referred to the discussion following eq.~\eqref{zetaperp}.]
The transverse dynamic MSD takes the form
\begin{equation}
  \delta r_\perp^2(t) \simeq \ell_\perp^3(t)/\ell_p \simeq
  (\Gamma_q^{(s)} t)^{3/4}/q^2,
\end{equation}
with $\Gamma_q^{(s)}\equiv(k_B T q^{8/3}/\zeta_\perp \ell_p^{1/3})$,
resulting in a stretched exponential decay of the dynamic structure factor, 
\begin{equation}
S(q,t) \propto \exp\left[-\frac{\Gamma(1/4)}{3 \pi} \left(\Gamma_q^{(s)} t\right)^{3/4}
\right],
\label{sqtstretchedexp}
\end{equation}
in agreement with earlier results
\cite{Farge1993,Kroy1997,Liverpool2001,Nyrkova2007}.
Equation \eqref{sqtstretchedexp} is a valid expression for times $t \ll \tau_e$
when the entanglement constraint is not felt. For simplicity we treated 
the value of a weakly time-dependent parameter arising in the
approximation to the hydrodynamic interactions as constant. Whenever the phenomenological
parameters need to be determined precisely from experimental data, the 
use of the improved expression for the MSD of a WLC of finite length
given in appendix \ref{hiapprox} is compulsory. A comparison of both results is provided
in Fig.~\ref{msdcomp}.

\paragraph{Terminal relaxation ($t \gg \tau_e$)}
For times longer than the `interaction' or entanglement time
$\tau_e\equiv (\zeta_\perp/\kappa) (L_e/\pi)^4$, defined here as the
relaxation time of a mode of (half) wavelength equal to the
characteristic interaction (or `entanglement') length $L_e$, the
decay of the structure factor slows down even more. This regime,
dominated by polymer-polymer interactions is observable in semidilute
solutions with $q L_e \approx 1$, where the scattering intensity can
still reasonably be described as an incoherent superposition of
single-polymer contributions but interactions slow down the relaxation
at long times. In most applications, interactions can in fact hardly
be avoided at concentrations yielding sufficient scattering
intensity. As derived in detail in section \ref{gwlc} and in appendix \ref{freevolume},
the GWLC predicts a crossover to a near power-law or
logarithmic relaxation, in accord with high precision dynamic
light scattering data \cite{Semmrich2007}. This differs from the
predictions of the dynamic tube model, reproduced in
eq.~\eqref{effmedium} below, which attributes the slowdown to
collisions of the polymer with a tube-like viscoelastic cage and
predicts an algebraic saturation of the MSD to a constant plateau
value.  While not unsuccessful in rationalizing experimental data
\cite{Liu2006,Vincent2007}, the tube model lacks the versatility to
account for the slanted plateaus ubiquitously observed.  In contrast,
the GWLC attributes the slowing down of the dynamics to an exponential
stretching of the relaxation spectrum characterized by a typical free
energy barrier height $k_B T \EE$ due to sticky interactions
with crossing polymers, or 
$k_B T \FF$ for purely steric interactions (cageing/entanglement).
We present results for the dynamic structure factor for the two limiting
cases, where only one of these contributions dominates.  
At long times, we write
\begin{equation}
\delta r_\perp^2(t \gg \tau_e) \approx\delta r_{\perp,L_e}^2(\infty) +
\delta r_{\perp}^{2,G}(t)
\label{rperpgwlc0}
\end{equation}
where $\delta r_{\perp,L_e}^2(\infty)=4 L_e^3/3 \ell_p \pi ^4$ is the contribution of the
non-interacting higher bending modes, i.e.\ the saturated static MSD of
a WLC of length $L_e$, and the MSD $\delta r_\perp^{2,G}(t)$ contains the slow dynamics
resulting from the stickiness or the steric interactions. With the help of
eq.~\eqref{sqtmsd}, the dynamic structure factor becomes 
\begin{equation}
\label{sfglassylongtime}
  S(q,t) \sim S(q,0) \exp[-q^2 \delta r_{\perp,L_e}^2(\infty)/4]G_{\EE,\FF}(q,t) \;,
\end{equation}
where 
\begin{equation}
G_{\EE,\FF}(q,t) \equiv \exp\left[-q^2 \delta r_\perp^{2,G}(t)/4 \right]
\label{sqtlongtime}
\end{equation}
The exponential factor in eq.~\eqref{sfglassylongtime} plays the role of
an Edwards-Anderson parameter in the limit of infinitely high free
energy barriers where $G_{\EE,\FF}(q,t)\rightarrow 1$.

\begin{figure}
\includegraphics[width=8cm]{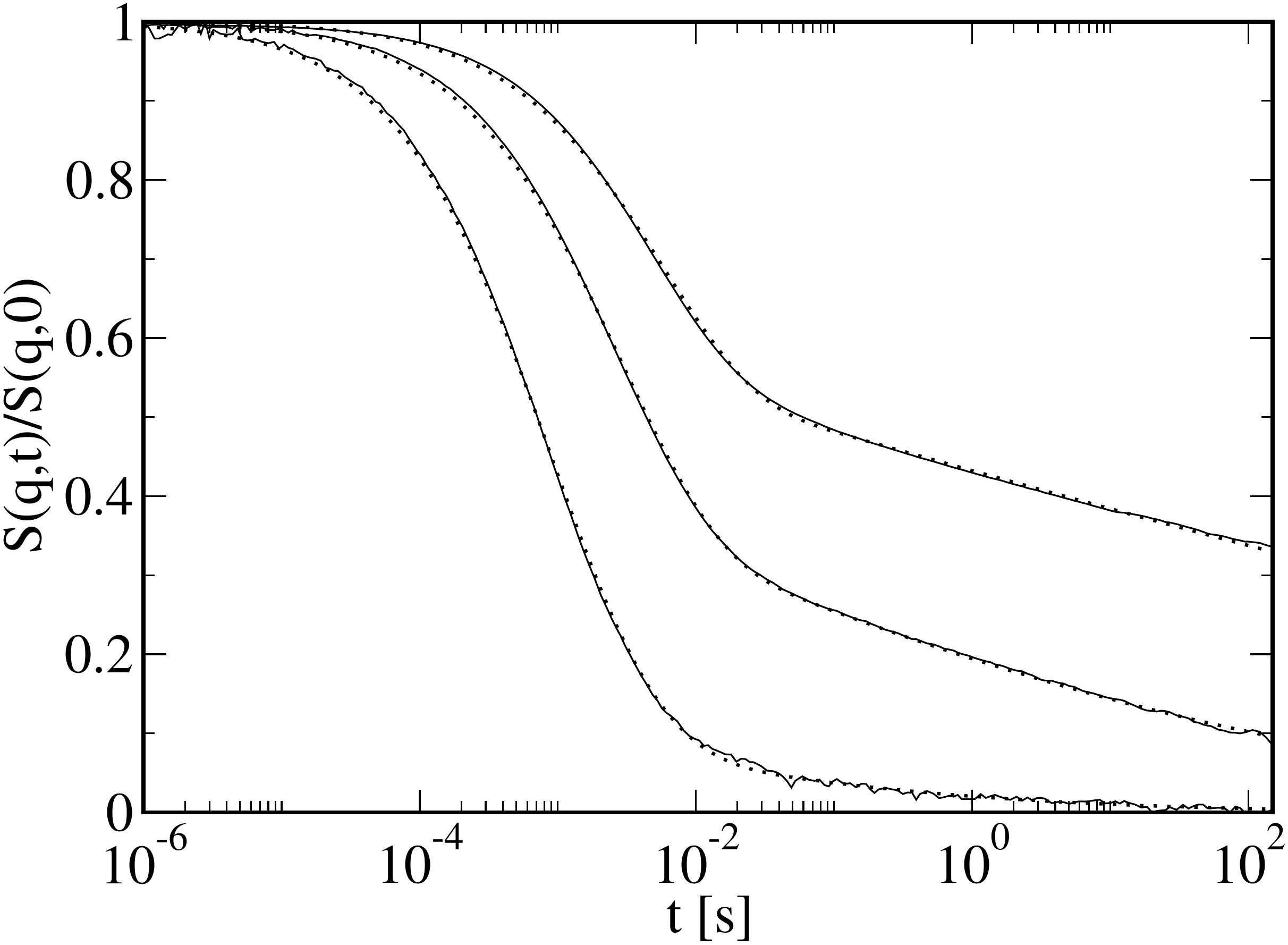}
\caption{Comparison of $S(q,t)$ to the experimentally determined dynamic structure
factor of (native) F-actin at temperature $T=15^\circ C$, actin concentration
$c=0.7\,\mbox{mg/ml}$.
Solid lines: DLS data for scattering vectors $q [\upmu \mbox{m}^{-1}]
=9.62, 17.13, 29.66$ (from top to bottom) \cite{Semmrich2007} (Original data
kindly provided by R. Merkel.)
Dotted lines: Dynamic structure factor of a GWLC, by numerical evaluation of eqs.
~\eqref{sqtmsd}, \eqref{msdint}. By using the analytical approximations,
eqs.~\eqref{sqtmsd}, \eqref{gwlclongtime}, \eqref{rperpgwlc} and \eqref{msdlogtot},
nearly indistinguishable fits can be obtained.
Values of the fit parameters: $\ell_p= 5.89\,\upmu\mbox{m}$ (determined
for $q=29.66\,\upmu\mbox{m}^{-1}$),
$\EE=43.40$ ($q [\upmu\mbox{m}^{-1}]=9.62$), $31.36$ ($q [\upmu\mbox{m}^{-1}]=17.13$).
For the highest $q$, $q^2 \Lambda^3/\ell_p \gg 1$.}
\label{sqtfit}
\end{figure}

The GWLC prediction for the MSD of a polymer with purely steric interactions
is derived in detail in appendix \ref{freevolume}, and the final asymptotic
result is:
\begin{multline}
\delta r_\perp^{2,G}(t) \approx \frac{4 L_e^3}{3 \ell_p \pi^4} \frac{1}{\FF^{3/4}}
\big\{\log^{3/4}\left[1+(t/\tau_e)\exp(\FF) \FF \right] \\
-\log^{3/4}\left[1+\exp(\FF) \FF\right] \big\} \qquad (t \FF e^{\FF}/\tau_e \rightarrow \infty,\,
 \FF \lesssim 1).
\label{msdglassyfreevolume}
\end{multline}
A result valid for all times is easily obtained with
the help of eq.~\eqref{rperpglassyfreevolume} below.
For long times the corresponding tail of the dynamic structure
factor $G_\FF(q,t)$ decays as a near power-law:
\begin{equation}
G_\FF(q,t) \propto t^{-\alpha(t)} \qquad (t \FF e^{\FF}/\tau_e \rightarrow \infty,\,
 \FF \lesssim 1),
\end{equation}
with a logarithmically time dependent exponent
\begin{equation}
\alpha(t) = \frac{q^2 \delta r_{\perp,L_e}^2(\infty)}{4 \FF^{3/4}}
\log^{-1/4}\left[(t/\tau_e)\exp(\FF)\FF\right].
\end{equation}
The GWLC prediction for the MSD of a sticky polymer at long times is given by
eq.~\eqref{msdint} below. To a good approximation valid for short and intermediate times
and for large $\EE$,
\begin{multline}
\delta r_\perp^{2,G}(t) \approx \frac{4 \Lambda^3}{\EE \ell_p \pi^4} [\gamma_E -
\mbox{Ei}(-t/\tau_\Lambda) + 
\log(t/\tau_\Lambda)] \\ (t/\tau_\Lambda \ll \EE, 1 \ll \EE).
\label{gwlclongtime}
\end{multline}
Here, $\mbox{Ei}$ is the exponential integral function.
The most salient feature of eq.~\eqref{gwlclongtime}
is an intermediate logarithmic relaxation that
becomes most pronounced for large $\EE\gg1$, which gives rise to a
logarithmic decay in the dynamic structure factor,
\begin{multline}
  G_\EE(q,t) \approx  1 - \frac{q^2 \Lambda^3}{\EE
      \ell_p \pi^4} \left(\gamma_E +
      \log\frac{t}{\tau_\Lambda}\right)\\\quad(\tau_\Lambda \ll t
  \ll \EE \tau_\Lambda)\;,
\label{sfglassylongtime2}
\end{multline}
where the arc-length distance $\Lambda$ between sticking sites can be identified with
$L_e$ for strong stickiness.
A direct numerical evaluation and a comparison with experimental data show that the
logarithmic tail in the
structure factor extends far beyond the upper bound on $t$ given in
eq.~\eqref{sfglassylongtime2} (compare Figs.~\ref{sqtfit},
\ref{fig:sqt_coll}).

It turns out that in many cases only one pair of new parameters ($\FF$
and $L_e$ or $\EE$ and $\Lambda$) is relevant: for purely repulsive interactions
$\Lambda \rightarrow \infty$,
while for (strongly) sticky interactions, $\Lambda \rightarrow L_e$ and
$\FF \rightarrow 0$.
A detailed experimental study of the crossover between the two idealized cases would
be desirable.

\section{The model --  equations of motion}

\subsection{The wormlike chain (WLC)}
We begin our formal discussion by introducing the WLC model and by
stating the equations of motion.  In the WLC model the polymer contour
is represented as a continuous space curve $\vec r(s)$ with a bending
energy
\begin{equation}
{\cal H}_{WLC}=\frac{\kappa}{2} \int \dd s\, \left(\frac{\partial^2
\vec r(s)}{\partial s^2} \right)^2.
\label{hwlc}
\end{equation}
The bending rigidity is denoted by $\kappa$. A key property of the WLC is its
inextensibility, expressed by the arc length constraint $|\vec
r'(s)|=1$. This nonlinear constraint renders the dynamical equations
of motion of the polymer difficult, but analytical progress can be
made in the weakly bending rod limit \cite{Hallatschek2007}, where the
polymer contour is parameterized by small transverse deflections
around the ground state, which is a straight line (chosen as the
$z$-axis). Introducing transverse and parallel coordinates,
\begin{equation}
\vec r(s,t)=[\vec r_\perp(s,t),s-r_\|(s,t)],
\end{equation}
the weakly bending approximation is formulated as a perturbation
calculation in the small parameter $\epsilon\equiv\overline{r_\|'}$,
where the over-bar denotes a spatial average. The arc length
constraint implies $r_\|'\approx\vec r_\perp'^2/2$. To lowest order, i.e.\  to order
$\epsilon^{1/2}$, there are no longitudinal fluctuations.

To specify the equations of motion, we need an expression for the viscous
drag. Hydrodynamic interactions are described with the help of the
hydrodynamic mobility tensor $\vec H(\vec r)$, which relates forces to
velocities. We use the following Rotne-Prager (RP) form
\cite{Dhont1996}:
\begin{multline}
\vec H(\vec r) = \frac{1}{3\pi \eta} \vec I \delta(r) + \frac{1}{8 \pi \eta}
\theta(r- d)(1/r) 
\big[{\vec I} (1+ d^2/6r^2) + \\
\vec {\hat r} \vec {\hat r} (1-d^2/2r^2) \big]
\end{multline}
Here, $d$ is the hydrodynamic diameter.  The first term accounts for
the Stokes friction of a monomer.  Neglecting non-diagonal terms
(which only contribute to higher order in $\epsilon$ in the transverse
equations of motion), we define the following mobility function for
the transverse undulations $\vec r_\perp(s,t)$:
\begin{equation}
h(s)=\delta(s)/3\pi \eta + \theta(|s|-d)(1+d^2/6s^2) /8 \pi \eta s.
\label{rptrans}
\end{equation}

The linear transverse equation of motion including hydrodynamic
interactions follows from eq.~\eqref{hwlc} as \cite{Kroy1997,Hallatschek2007,Nyrkova2007}
\begin{multline}
\partial_t \vec r_\perp(s,t) = \int\dd s'\, h(s-s')
 [-\kappa \vec r_\perp''''(s',t) +
\xiv_\perp(s',t)],
\label{eomtrans}
\end{multline}
where $\xiv_\perp$ is the transverse Gaussian white noise. 

The transverse equation of motion is solved by introducing
normal modes. For simplicity and without serious consequences for our results
we impose hinged ends\footnote{For free boundary conditions, we expect
some influence of the different terminal relaxation times and amplitudes resulting from
different values of the low wave numbers $k_n$ \cite{LeGoff2002}.
Our discussion is therefore restricted to times shorter than the terminal relaxation time
$\tau_L$, defined below eq.~\eqref{msdfull}.}
as boundary conditions,
\begin{equation}
\vec r_\perp(s,t) = \sqrt{\dfrac{2}{L}} \sum\limits_{n=1}^\infty \vec a_n(t) \sin(k_n s),
\label{rperpmode}
\end{equation}
where $k_n=n \pi/L$ are the wave numbers.
The equation of motion eq.~\eqref{eomtrans} for
$f=0$ is rewritten in mode space as
\begin{equation}
\partial_t \vec a_n(t) = \sum\limits_{nm} h_{nm} [-\kappa k_m^4 \vec a_m(t)
+ \xiv_m(t)],
\label{eommodespace}
\end{equation}
where we have introduced the mobility matrix
\[
h_{nm} \equiv \frac2L \int\dd s\,\dd s'\, \sin(k_n s) \sin(k_m s')
h(s-s') \;.
\]
For $n,m\gg 1$ it reduces to $h_{nm}=\delta_{nm} \tilde h(k_n)$, where
the mode mobility $\tilde h(k)$ is the Fourier transform of $h(s)$
\cite{Doi1988}.  For the Rotne-Prager form of the mobility, we thus
obtain the following approximation to the mode friction (for $kd \ll 1$):
\begin{equation} 
  \zeta_\perp \equiv 1/\tilde h(k) \approx4 \pi
  \eta/[C'-\log(kd)]\;.
\label{zetaperpk}
\end{equation}
The constant $C'$, which takes the value $C'=17/12 - \gamma_E \approx 0.84$, is
characteristic of the particular form of the hydrodynamic interactions
chosen.  On this level of description, the hydrodynamic interactions
are completely encoded into this mode dependent friction coefficient
of the independent normal modes. As usual, the modes relax
individually and exponentially,
\begin{equation}
C_{nm}(t) \equiv \langle \vec a_n(t) \vec a_m(0)  \rangle = \delta_{nm} \langle
\vec a_n^2 \rangle
\exp(-t/\tau_n)\;.
\end{equation}
Here we defined the relaxation time of mode number $n$, $\tau_n \equiv
(\zeta_\perp/\kappa) k_n^{-4}$. The equilibrium mode amplitudes follow
from the equipartition theorem as $\langle \vec a_n^2 \rangle =
2/\ell_p k_n^4$.  

\begin{figure}
\centering\includegraphics[width=8cm]{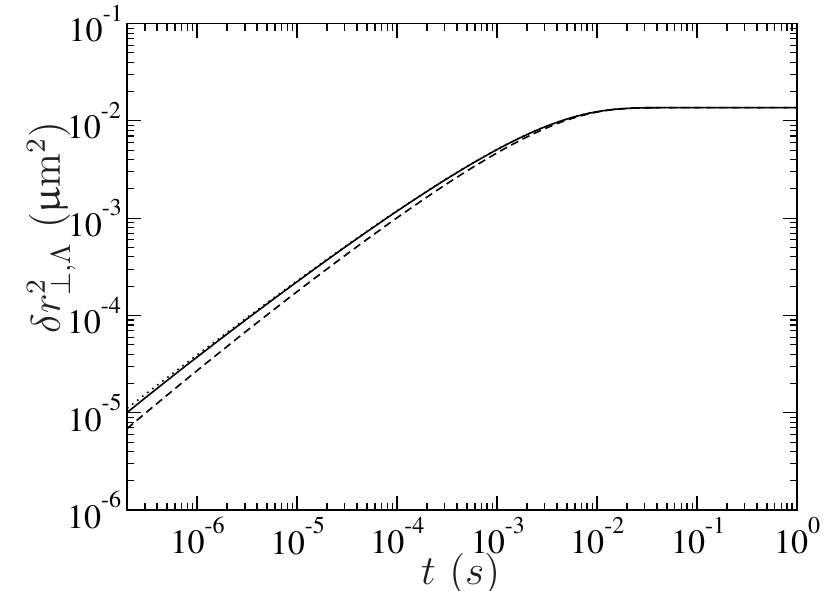}
\caption{Comparison of different approximations of the MSD $\delta r_{\perp,\Lambda}^2$
of a WLC of length $\Lambda=1\,\upmu\mbox{m}$ ($\ell_p=1\, \upmu\mbox{m}$) with
hydrodynamic interactions: numerical evaluation of eq.~\eqref{msdlog} (---),
simple analytical approximation eq. ~\eqref{rperplambda} with time dependent
friction constant $\zeta_\perp(t)$ ($-\,-$) and improved
approximation eq.~\eqref{msdlogtot} ($\dots$).}
\label{msdcomp}
\end{figure}

The most important observable for our discussion of the dynamic
structure factor is the dynamic part of the MSD
\begin{equation}
  \delta r_\perp^2(s,s',t) \equiv 2[\langle \vec r_\perp(s,0) \vec
  r_\perp(s',0)  \rangle
  - \langle \vec r_\perp(s,t) \vec r_\perp(s,0) \rangle ]\;
\end{equation}
of contour element $s$ with respect to contour element $s'$.
In mode space, this takes the form
\begin{multline}
  \delta r_\perp^2(s,s',t) = \frac{4}{L \ell_p} \sum\limits_n
  \left\{\cos[k_n (s'-s)]
    +\cos[k_n (s+s')] \right\}\\
  \frac{1-\exp(-t n^4/\tau_L)}{k_n^4},
\label{msdfull}
\end{multline}
where $\tau_L\equiv \tau_1=(\zeta_\perp/\kappa) (L/\pi)^4$. This
equation may be further simplified by averaging over the variable $s$
with $s'-s=\mbox{const.}\ll L$ (denoted by an over-bar), which is a valid
procedure everywhere except in vicinity of size $\ell_\perp(t)$ (to be defined
below) of the ends,
\begin{equation}
  \overline{\delta r_\perp^2(s,s',t)} = \frac{4}{L \ell_p} \sum\limits_n 
  \cos[k_n (s'-s)]
  \frac{1-\exp(-t n^4/\tau_L)}{k_n^4}\;.
\label{msdsum}
\end{equation}
In the following we evaluate the above sum on two different levels of approximation.
We first present a simplified discussion, exactly valid in the limit
$L \rightarrow \infty$. In appendix \ref{hiapprox} an improved approximation 
is discussed, and the expression given there should be preferred over the
following eqs.~\eqref{dynmsd},\eqref{zetaperp} --- or their analogue for a WLC of finite
length, eq.~\eqref{rperplambda} --- for quantitative purposes.
A comparison of both approximations is shown in Fig.~\ref{msdcomp}.

We now proceed with the simpler, but less accurate expression for an infinite chain.
The sum eq.~\eqref{msdsum} is converted into an integral, and
after a change of variables
$z=\ell_\perp(t) k_n$,
\begin{multline}
  \overline{\delta r_\perp^2(s,s',t)} = \frac{4}{\pi}
  \frac{\ell_\perp^3(t)}{\ell_p} \int\limits_0^{\ell_\perp(t)/d} \dd
  z\,  \cos\left[\frac{z (s-s')}{\ell_\perp(t)}\right] \\
  \times\frac{1-\exp(-z^4)}{z^4}\;,
\label{dynmsd}
\end{multline}
where we assumed that $\ell_\perp(t) \ll L$. Here, the weakly varying logarithm in the
mode friction
$\zeta_\perp(k)$ is treated as a (time dependent) constant with respect to $k$,
\begin{equation}
\zeta_\perp(t)\approx \frac{4 \pi \eta}{C'-\log[d/\overline\ell_\perp(t)]},
\label{zetaperp}
\end{equation} 
We have also introduced the two abbreviations
$\ell_\perp(t)$ and $\overline\ell_\perp(t)$ for the (exact and
approximate) transverse correlation length $\ell_\perp(t) \equiv
(\kappa t/\zeta_\perp)^{1/4}$ and $\overline\ell_\perp(t) \equiv
(\kappa t/4 \pi \eta z^\star)^{1/4}$, respectively,
with $z^\star$ being a weakly time-dependent effective mode number
of order unity.
For qualitative purposes it is sufficient to substitute $\overline\ell_\perp(t)
\rightarrow q^{-2/3}\ell_p^{1/3}$ in the argument of the logarithm of eq.~\eqref{zetaperp},
which yields a simple expression for $\zeta_\perp(t)$
\cite{Nyrkova2007}. 

If the $z$-integration in eq.~\eqref{dynmsd} is carried out for $s=s'$,
the dynamic MSD $\delta r_\perp^2(t)\equiv \overline{\delta r_\perp^2(s,s,t)}$ of a
free polymer of infinite length is recovered:
\begin{equation}
\delta r_\perp^2(t)=\frac{\ell_\perp^3(t)}{\pi \ell_p} \frac{\Gamma(1/4)}{3}
= \frac{\Gamma(1/4)}{3\pi q^2} (\Gamma_q^{(s)} t)^{3/4}.
\end{equation}

The full mean square displacement 
consists of a static and the dynamic part,
\begin{equation}
\begin{split}
\Delta r_\perp^2(s,s',t) &\equiv \langle [\vec r_\perp(s,t) - \vec r_\perp(s',0)]^2 \rangle
\\
&= \langle [\vec r_\perp(s,0)-\vec r_\perp(s',0)]^2 \rangle + \delta r_\perp^2(s,s',t)
\end{split}
\end{equation}
It is easy to see by a Taylor expansion, that in the bulk of the
polymer the static MSD between two different points on the contour is
to leading order quadratic in the contour length, $\Delta
r_\perp^2(s,s',0) \simeq \epsilon (s-s')^2/2$.
More precisely, a systematic calculation
for hinged ends gives \cite{glaser:unpub2}
\begin{multline}
\Delta r_\perp^2(s,s')=
  \frac{1}{6} \ell_p^{-1} L^3 \Big\{ (\tilde s'-\tilde s)^2 \big\{1+ \\
  4 \big[(1/2-\tilde s)^2 +
  (1/2-\tilde s)(1/2-\tilde s')+(1/2-\tilde s')^2\big]\big\}\\
  - 2 |\tilde s'-\tilde s|^3 \Big\} \quad
  (\tilde s\equiv s/L,\,\tilde s'\equiv s'/L).
\end{multline}
The terms in the middle as well as the last one can be neglected in the bulk of the
polymer, where $|s-L/2| \ll L/2$.
For long times, only the 
dynamic part of the MSD evaluated at $s=s'$ contributes to the deacy of the
dynamic structure factor, resulting in the incoherent dynamic structure factor
\cite{Kroy2000}.

In the remainder of this section, we discuss extensions of the theory
for isolated filaments that address the effects of the surrounding
solution.
For completeness, we first briefly summarize the idea behind
and the most important prediction of the standard tube model, before
we describe the alternative GWLC model, which compares more favorably
to a large body of experimental data.

\subsection{The tube model}
Quasi-static quantities, such as the plateau modulus of a solution of
semiflexible polymers are satisfactorily explained by the tube model,
which assumes that the effect of the surrounding network may
effectively be represented by a tube-like cage, see
e.g. \cite{Isambert1996,Morse2001,Kroy2006,Hinsch2007}.  The tube
properties are characterized by the entanglement length $L_e$ (the arc
length over which the bending energy equals the confinement energy) or
equivalently by the tube diameter $d$. Both quantities are related by
the basic roughness relation $d^2\simeq L_e^3/\ell_p$ of the WLC,
which also implies that the transverse mean-square deflections from a
clamped end grow like the third power of the wavelength.

In the simplest static version of a harmonic tube potential, which is
added to the Hamiltonian, the relaxation of a confined polymer is
exponentially suppressed for times longer than the entanglement time
$\tau_e$ associated with a mode of wavelength $k^{-1} \simeq L_e$
\cite{Janmey1994}. The saturation of the MSD and the dynamic structure
factor to their plateau values is however too quick if compared to
experimental observations.  This is improved by an extension of the
tube model that includes dynamic fluctuations of the tube arising from
an over-damped homogeneous elastic background material
\cite{Kroy2000}, which yields
\begin{multline}
\delta r_\perp^2(t)=\langle \vec r^2 \rangle \left[1-\frac{\sqrt{\pi}}{2}
\frac{\mbox{erf}(\sqrt{\omega_\star t})}{\sqrt{\omega_\star t}}\right] +\\
\langle \vec r_\perp^2 \rangle
\left[1- \frac{3}{4} (\omega_\star t)^{3/4} \Gamma(-3/4,\omega_\star
  t)\right] \;.
\label{effmedium}
\end{multline}
Here, the prefactors $\langle \vec r^2 \rangle$ and $\langle \vec
r_\perp^2 \rangle$ are the mean squared amplitudes of the effective
medium and of the polymer, respectively, $\omega_\star$ is
approximately (but not strictly) identical to the inverse of the
entanglement time $\tau_e$. Theoretically, the static MSD, to which
the dynamic MSD saturates algebraically (like $t^{-1/2}$), is
connected to the crossover frequency $\omega_\star$ via
\begin{equation}
\langle \vec r_\perp^2 \rangle = \frac{4}{3 \pi} \left(
\frac{k_BT \omega_\star^{-1}}{\zeta_\perp \ell_p^{1/3}} \right)^{3/4}\;.
\end{equation}
While this model seems to agree reasonably well with experimental data
\cite{Mason2000,Liu2006,Vincent2007} if the amplitudes and the
crossover frequency are treated as free fit parameters, it cannot
account for the `slanted plateaus' generally observed, which
correspond to a very slow terminal relaxation of the MSD.

\subsection{The basic idea of the glassy wormlike chain (GWLC)}
It is only very recently that high-precision DLS experiments
unambiguously demonstrated a slow logarithmic terminal relaxation
instead of a plateau in the dynamic structure factor of actin
solutions at low temperatures \cite{Semmrich2007}.  This has
been interpreted as a signature of a system near its glass transition,
which only slowly relaxes into equilibrium.  An interpretation of the
logarithmic tails in the framework of the established mode coupling
theory for glasses \cite{Gotze1991}, as suggested by recent simulation
studies of flexible polymer blends \cite{Moreno2006}, would require an
improbable fine tuning of parameters into the neighborhood of a
higher-order mode coupling singularity. It therefore seems at variance with
the generic nature of the slow relaxation in actin solutions, which is
found to extend over more than three decades in time
\cite{Semmrich2007} and observed for a wide range of
concentrations. For the same reasons, and additionally for the lack of
any observations of a percolation structure in actin solutions, an
interpretation in terms of a percolation critical point seems
unnatural to us.

\begin{figure}
\center \includegraphics[width=8cm]{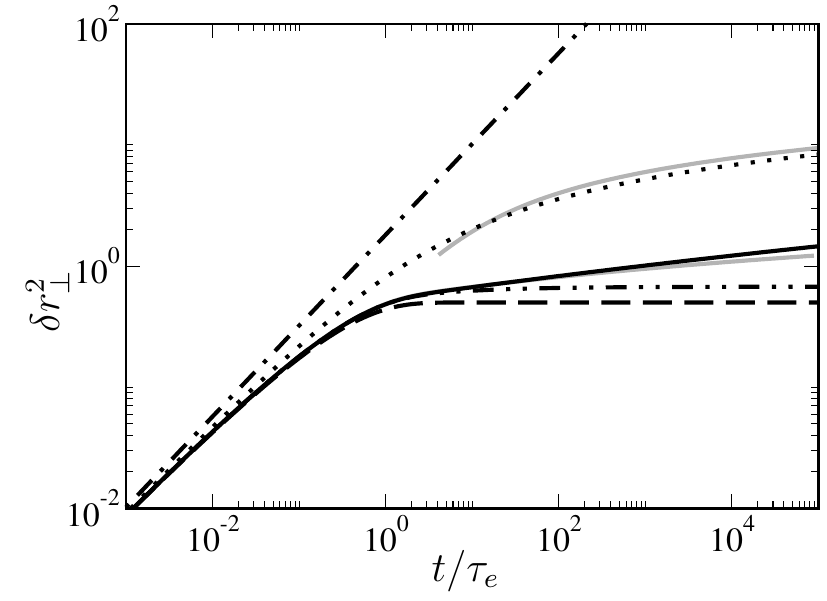}
\caption{MSD of a free polymer ($-\cdot-$), of a polymer in
an overdamped elastic background (--~$\cdot\cdot$) 
with $\langle \vec r^2\rangle =0.35\, \langle \vec r_\perp^2\rangle$,
GWLC with purely steric interactions ($\cdots$), and a GWLC with purely sticky
interactions according to $\Lambda=L_e$, $\EE=\infty$ ($- -$) (equivalent to a static
tube) and $\EE=25$ (---), all with $\delta r_{\perp,L_e}^2(\infty)=\langle
\vec r_\perp^2 \rangle/3 =4 L_e^3/3\ell_p \pi^4=0.5$. The approximate analytical
expressions eqs.~\eqref{gwlclongtime}, \eqref{rperplambda} and the asymptotic result
eqs.~\eqref{msdglassyfreevolume}, \eqref{rperplambda} (for
$t > \tau_e e^{-\FF}\!/\FF$ with $\Lambda \rightarrow L_e$)
are shown in gray.}
\label{fig:msd_gwlc_tube}
\end{figure}

In contrast to models attributing the stretching of the relaxation
spectrum to a critical point, the GWLC \cite{Kroy2007}
suggests a very intuitive origin of the slow-down in a rough free
energy landscape.  The basic assumption of the model in its simplest
form is that short wavelength modes with a (half) wavelength
$\lambda_n \equiv \pi/k_n <\Lambda$ can relax freely, while for the
relaxation of a mode $n$ with $\lambda_n>\Lambda$ a certain number
$N_n$ of energy barriers of height $\EE k_BT$ have to be overcome.
This prescription is in the spirit of the generic trap models recently
favored by many experimental investigators of cell mechanics
\cite{Fabry2001,Lenormand2006}, but it puts them on a more concrete
basis.  Physically the parameter $\EE$ is thought to arise primarily
from direct adhesive polymer interactions \cite{Kroy2007}.
But in the same vein steric caging and entanglement effects may be cast into the
language of free energy barriers by a free volume argument.
Assigning an Arrhenius factor for each free energy
barrier, the above prescription gives rise to an anomalous stretching
of the WLC relaxation spectrum that manifests itself in anomalously
slow (logarithmic) decay of single-polymer conformational correlations
at long times.  Implementing the corresponding prescriptions in
eq.~\eqref{msdint} for the dynamic MSD of a WLC, one obtains
the predictions of the GWLC for the late time
dynamics. Fig. \ref{fig:msd_gwlc_tube} illustrates the effect of
different
types of interactions of the test chain with its surrounding medium on
the dynamic MSD, and compare them to the predictions of the classical
tube model. A similar comparison for the structure factor is presented
further below.
We remark that experimental evidence
of logarithmic relaxation (or near-constant loss in the susceptibility spectra)
observed for synthetic polymer melts \cite{Sokolov2001} could be indicative of a
mechanism akin to that proposed for the GWLC in these systems. This observation has
recently prompted the introduction of the `glassy Rouse chain'
\cite{Glaser2008}.

\subsection{Direct sticky interactions in the GWLC}
\label{gwlc}

In the simplest version, the GWLC behavior arises from very short-ranged (sticky)
interactions of crossing polymers, as they might arise from hydrogen bonds or hydrophobic
patches on a biopolymer's backbone, or dispersion forces that
are cut off at short distances. If such interactions are imperfectly
screened by electrostatic
repulsions, the resulting pair interaction potential will feature an
energy barrier, which is denoted by $\EE k_BT$
\cite{Kroy2007}. While such interactions may under some
conditions induce phase separation, we concentrate on conditions of moderate
attractions, where they have no thermodynamic but merely a \emph{kinetic} effect via
the barrier. In
this case, the strategy of keeping the equilibrium mode amplitudes
$\langle \vec a_n^2\rangle$ of the WLC unchanged while modifying its
relaxation spectrum should be adequate. (In fact, the question how the
equilibrium amplitudes are renormalized by the presence of a
disordered background poses a formidable theoretical challenge, but
the following analysis suggests that it might not be the dominant effect for
the dynamics.)
 
Whenever such sticky interactions are the dominant effect, the
relaxation spectrum of the bare WLC is modified according to
\begin{equation}
\tilde\tau_n =  \begin{cases}
\tau_n & n>n_\Lambda = L/\Lambda\\
\tau_n \exp(N_n \EE ) &n<n_\Lambda\\
\end{cases}
\label{newtaun}
\end{equation}
with
\begin{equation}
  N_n\equiv(\lambda_n/\Lambda - 1)\;.
\end{equation}
Here, $n_\Lambda$ denotes the interaction wave number, so
that only two new parameters are introduced: the stretching parameter
or barrier height $\EE$ and the interaction length $\Lambda$, which has the direct
physical interpretation as a typical contour distance between sticky contacts.
For example, if the polymer contour as a whole is highly sticky,
$\Lambda$ will be identical to the entanglement length $L_e$. When the
stickiness is only induced at certain places along the backbone, e.g.\
if it is due to an incomplete coverage of the backbone with some
molecular crosslinker, $\Lambda$ may be substantially larger than
$L_e$. The same role is actually played by the strength of the
attraction in the sticky case, where it determines the equilibrium ratio of bound to
unbound entanglement points and thus $\Lambda/L_e$ via a Boltzmann factor. 
In particular in cases where $\EE\lesssim 1$ and/or
$\Lambda\gg L_e$ it becomes important to also consider the effect of steric
interactions between adjacent polymers, known as cageing or
entanglement, which is the subject of the following paragraph.

\subsection{Cageing and entanglements in the GWLC}
The GWLC model as introduced above is applicable to systems dominated
by adhesive interactions.  Additional contributions to the stretching
of the relaxation spectrum arise however from steric interactions of the chain with its
surroundings. Our basic assumption is that for each wavelength there are several
substantially different conformations of similar free energy available. 
In order to relax the caged long wavelength modes, the
surrounding matrix then has to be temporarily pushed out of the way to create enough free
volume for the conformational change, a process that can again be described by an escape
over a free energy barrier. More precisely, a wavelength dependent 
entanglement volume $V_\lambda$ may be defined 
as the free volume needed by a mode of (half) wavelength $\lambda$ to
relax, which depends on the magnitude of the transverse excursions. These may be inferred
from the following simple argument: With the help of eq.~\eqref{rperpmode} the
MSD of a monomer after relaxation of a mode of wavelength $\lambda=L/n$ is calculated as
\begin{equation}
\delta r_{\perp,\lambda}^2 = \frac{2}{L} \sum\limits_{n=L/\lambda}^\infty \langle \vec a_n^2
\rangle \approx \frac{4 L^3}{\pi^4 \ell_p} \int\limits_{L/\lambda}^\infty
\frac{\dd n}{n^4}\, \simeq \lambda^3/\ell_p
\end{equation}
(where the spatial averaging has already been carried out).
Accordingly, we have $V_\lambda = \lambda \delta
r_{\perp,\lambda}^2 \propto \lambda^4$.
The free energy barrier, that a mode of (half) wavelength $\lambda$ has to
overcome to relax in presence of a background polymer solution of entanglement length $L_e$
is then estimated as $k_B T \FF \lambda^4/L_e^4$. Moreover, it is
assumed that modes with a transverse MSD smaller than
$\delta r^2_{\perp,L_e}$ can relax freely. Similar to the above case with
adhesive interactions, this leads to an additional stretching of the
relaxation spectrum, where the relaxation times of modes of wavelength
$\lambda_n > L_e$ are slowed down according to
\begin{equation}
  \tilde\tau_n =  \begin{cases}
    \tau_n & n>n_e=L/L_e \\
    \tau_n \exp(N_n' \FF ) &n<n_e\\
\end{cases}
\label{taunfreevolume}
\end{equation}
with
\begin{equation}
  N_n'\equiv(\lambda^4_n/L_e^4 - 1).
\end{equation}

\begin{figure}
\center \includegraphics[width=8cm]{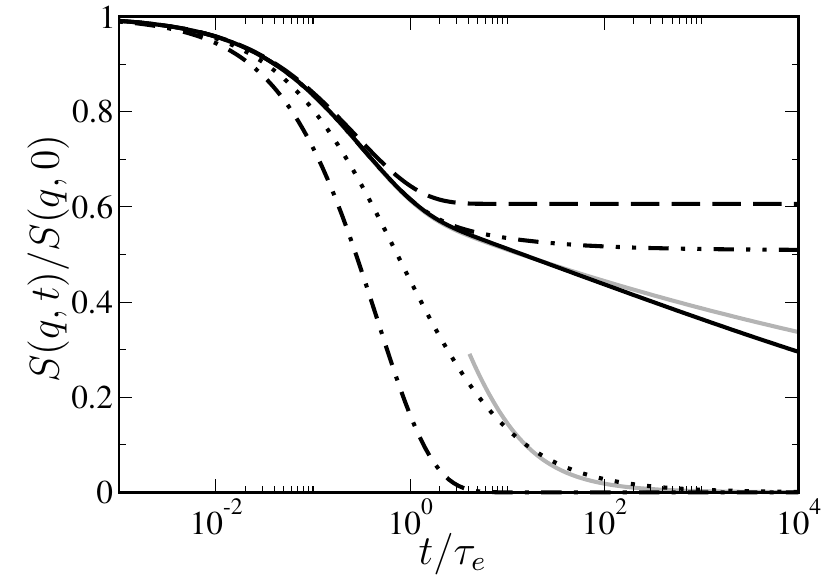}
\caption{Dynamic structure factor of a free polymer ($-\cdot-$), of a
polymer in an overdamped elastic background (--~$\cdot\cdot$) 
with $\langle \vec r^2\rangle =0.35\,\langle \vec r_\perp^2 \rangle$, of a GWLC
with purely
steric interactions ($\cdots$), and a GWLC with purely sticky interactions according to
$\Lambda=L_e$ and $\EE=\infty$ ($- -$) (equivalent to a static tube) and
$\EE=25$ (---), all with
$\delta r_{\perp,L_e}^2(\infty)=\langle \vec r_\perp^2 \rangle/3 
=4 L_e^3/3\ell_p
\pi^4=0.5$.
The dynamic structure factors obtained from the approximate analytical expressions
eqs.~\eqref{sqtmsd}, \eqref{gwlclongtime}, \eqref{rperplambda}
and the asymptotic result eqs.~\eqref{sqtmsd},
\eqref{msdglassyfreevolume}, \eqref{rperplambda} (for $t > \tau_e e^{-\FF}\!/\FF$ 
with $\Lambda \rightarrow L_e$) are shown in gray.}
\label{fig:sqt_coll}
\end{figure}

The dimensionless free energy barrier height $\FF$ is generally expected to have a
small numerical value, as can be seen from the following
argument. The interaction free energy of a rigid polymer segment with hard core
interactions is estimated as $c v k_B T/2$, where $c$ is the concentration of collision
points and $v$ is the excluded volume of the segments.
Here we consider semiflexible polymers of length $L_e$ with purely hard core direct
interactions that are represented as soft
cylinders interacting via an effective potential $V(r)$ on a coarse-grained level.
Comparing their mutual
repulsion to a hard core interaction, we estimate $\FF \simeq 2 B_2/ v$, where $B_2$ is the
second virial coefficient of the solution of soft cylinders.
The effective interaction potential between the cylinders,
is due to a kind of Helfrich repulsion \cite{Helfrich1985,Daniels2004}.
Consider two WLCs of length $L_e$ that approach each other
at orthogonal orientations.  If the two axes through the polymer endpoints are held at a
fixed distance $r$ from each other, their effective interaction potential $V(r)$ is found
to be \cite{Teeffelen,Morse2001}
\begin{equation}
  V(r)=-k_B T
  \log\left[\mbox{erfc}(-2 \sqrt{3} r/d)/2\right] \;,\qquad d=\sqrt{L_e^3/\ell_p},
\end{equation}
From this, the second virial coefficient is calculated as an integral
over the Mayer $f$-function,
\begin{equation}
  B_2 = - \frac{1}{2} \int\dd\,\vec r \int \frac{\dd \Omega}{4 \pi}\,
  \int\frac{\dd \Omega}{4 \pi}'\,  \{\exp[-\beta V(\vec r,\Omega,\Omega')]-1\},
\end{equation}
where $\vec r$ denotes relative separation between
the two axes and $\Omega,\Omega'$ denote their orientations. The implicit
assumption is made that the potential of the two polymers crossing at an arbitrary
angle is still of the form for orthogonally crossing polymers. The
integral is split into a surface integral over a parallelogram (with edges
of the tube length $L_e$) in the plane spanned by two unit vectors in the direction of
$\Omega, \Omega'$ (the excluded area) and an integral over the distance of the cylinders
in a direction transverse to this plane, and it is assumed that the
interaction potential only depends on this distance. 
One arrives at $B_2=L_e^2 d \sqrt{\pi}/8 \sqrt{3}$, and using $v= \pi L_e^2 d/2$ for
the excluded volume of hard cylinders of diameter $d$ and length $L_e$
\cite{Onsager1949}, we get:

\begin{equation}
  \FF \approx \frac{1}{\sqrt{\pi} 2 \sqrt{3}} \approx 0.2.
\label{estf}
\end{equation}
For the strength of the steric interactions in 
Figs.~\ref{fig:msd_gwlc_tube}-\ref{fig:sqt_coll}, this value of $\FF$ is used. It should
be noted that $\FF$ must in principle be recalculated if attractive interactions are
present, in which case its absolute value can be substantially smaller. For strongly
attractive interactions, $\FF$ may turn negative, indicating that the free volume approach
breaks down.
The value for $\FF$ obtained in eq.~\eqref{estf} should thus be understood as a rough
estimate of an upper limit for purely steric interactions, while in the case
$\Lambda \simeq L_e$ of strong stickiness, $\FF \rightarrow 0$.

\section{The dynamic structure factor}
\label{resultslong}

\subsection{Initial decay ($t \ll \tau_q^\perp$)}
In this section we discuss the initial decay regime of the dynamic
structure factor of (isolated) semiflexible filaments. Our calculation
will be different from that in refs.
\cite{Kroy1997,Liverpool2001,Nyrkova2007}, which was performed in mode
space or Fourier space respectively, whereas our approach employs a
real space representation.

To perform the thermal average in the dynamic structure factor
eq.~\eqref{cohsf}, the assumption that $\vec r_\perp$ is a Gaussian distributed variable
is employed, valid for situations where the scattering can be traced back to
single polymers which are described by eq.~\eqref{eomtrans}, i.e., whenever the scattering
wavelength is smaller than the mesh size ($q^{-1} \ll \xi$). A quantitative evaluation of
the derived results should thus focus on the largest measurable values of $q$.
Neglecting longitudinal fluctuations, we get
\begin{multline}
S(q,t)=\frac{1}{Ld}  \int\dd s\int\dd s' \langle\exp[-q_\perp^2
\Delta r_\perp^2(s,s',t)/4\\+
i q_\| (s - s')] \rangle_O,
\label{sfexp}
\end{multline}
where $q_\|=q \cos(\theta)$ and $q_\perp^2=q^2 \sin^2(\theta)$.
\begin{equation}
\langle \dots \rangle_O\equiv (1/2) \int\limits_0^\pi \dd \theta\,\sin(\theta)
\end{equation}
denotes an average over orientations of the polymer. Taking this
average, we get in the limit $q^2 \Delta r_\perp^2(t) \ll 1$ ($t
\rightarrow 0$)
\begin{multline}
  S(q,t) \approx \frac{1}{2Ld} \int\dd s\,\dd s'\,\int\limits_{-1}^1 \dd x\,
  [1-q^2 (1-x^2) \Delta r_\perp^2(s,s',t)/4\\ + \dots] \exp[i q x
  (s-s')].
\label{sfseries}
\end{multline}
For $t=0$ and to zeroth order in $\epsilon$, the expression for the
static structure factor is obtained:
\begin{equation}
  S(q,0) \approx \frac{1}{2 L d} \int\dd s\,\dd s'\, \int\limits_{-1}^1\dd x\,
  \exp[iqx(s-s')] = \frac{\pi}{q d}.
\label{staticsf}
\end{equation}
The initial decay rate is defined as
\begin{equation}
\Gamma_q^{(0)}\equiv-\lim_{t\rightarrow 0}\frac{\dd}{\dd t} \log(S(q,t)).
\end{equation}
We therefore need to calculate $\partial_t \Delta
r_\perp^2(s,s',t)$. We have
\begin{multline}
  \lim_{t\rightarrow 0} \partial_t \Delta r_\perp^2(s,s',t) =
  -2 \lim_{t\rightarrow 0} \langle \vec r_\perp(s',0) \partial_t \vec r_\perp(s,t)\rangle\\
  +2 \lim_{t\rightarrow 0} \langle \vec r_\perp(s,t) \partial_t \vec
  r_\perp(s,t)\rangle.
\label{corr1}
\end{multline}
The last term in eq.~\eqref{corr1} is actually a time derivative of an
equilibrium correlation function and vanishes consequently. We replace
the first time derivative using the equation of motion,
eq.~\eqref{eomtrans}, and thus get an integral over an equilibrium
correlation function for the transverse coordinates:
\begin{multline}
\lim_{t\rightarrow 0} \partial_t \Delta r_\perp^2(s,s',t)\\
=2 \kappa \lim_{t\rightarrow 0}
\int_0^L\dd\tilde s\,h(s-\tilde s) \langle \vec r_\perp''''(\tilde s,t)
\vec r_\perp(s',0) \rangle
\label{dtdeltar}
\end{multline}
The last correlator is obtained via the equipartition theorem, which
yields
\begin{equation}
\langle \vec r_\perp''''(\tilde s) \vec r_\perp(s') \rangle = 2 \ell_p^{-1}
\delta(\tilde s-s'),
\label{equipartition}
\end{equation}
taking into account two transverse directions.  Combining
eqs.~\eqref{sfseries}, \eqref{dtdeltar} and \eqref{equipartition} we
get
\begin{multline}
  \lim_{t\rightarrow 0}\partial_t S(q,t) = - \frac{k_B T q^2}{2 L d} \int\dd s\,\dd
  s'\, \int\limits_{-1}^1\dd x\, h(s-s') (1-x^2)\\ \times\cos(q x
  (s-s')).
\end{multline}
With the help of the general identity
\begin{equation}
\int\limits_0^L \dd s\,\int\limits_0^L \dd s'\, f(s-s') =
\int\limits_{-L}^L\dd s\, f(s) (L - |s|)
\end{equation}
and eq.~\eqref{staticsf} we find the general expression for the
initial decay rate
\begin{align}
  \Gamma_q^{(0)}&= \lim_{t\rightarrow 0}\frac{\partial_t S(q,t)}{S(q,0)}\\
  &=\frac{k_BT q^3}{2 \pi L} \int\limits_{-L}^L\dd
  s\,\int\limits_{-1}^1\dd x\, (L-|s|) h(s) (1-x^2) \cos(q x s).
\label{inidecgeneral}
\end{align}

Using the mobility function, eq.~\eqref{rptrans}, an evaluation of
eq.~\eqref{inidecgeneral} for $d \ll q^{-1} \ll L$, an assumption
usually fulfilled in light scattering experiments on biopolymers,
gives
\begin{equation}
\Gamma_{q,RP}^{(0)} = \frac{k_B T q^3}{6 \pi^2 \eta} [C-\log(qd)].
\label{inidecshort}
\end{equation}
(terms of order $(qd)^2$ and higher have been discarded and the limit
$L\rightarrow \infty$ has been taken.)  The constant is
$C=4/3+17/12-\gamma_E=2.17$, where $\gamma_E$ is Euler's constant.
This value differs from the previously reported value of $C=5/6$
\cite{Kroy1997}.  We attribute this to our improved treatment of the
hydrodynamic interactions.  However, it also shows that the exact
value sensitively depends on the details of the hydrodynamic model,
which has to be taken into account for the determination of the
hydrodynamic backbone diameter from measurements of the initial decay
rate.  This is also acknowledged in \cite{Nyrkova2007}. While our
result confirms the previous results found in
\cite{Kroy1997,Nyrkova2007}, Liverpool and Maggs \cite{Liverpool2001}
obtained a qualitatively different result, where an additional
logarithmic factor $S(q,t)/S(q,0) -1 \simeq t \log(t)$ appears in the
short time behavior of the dynamic structure factor, diverging in the
limit $t\rightarrow 0$. This results from their too inaccurate
treatment of the spherical Bessel functions occurring in the discussion
in mode space.

We remark that if one chooses the simpler Oseen expression for the
mobility function, our result for the initial decay rate due to transverse
displacements, eq.~\eqref{inidecgeneral}, reduces (after carrying out the integration
over $x$ and in the limit $L\rightarrow \infty$) to a result quoted in
refs.~\cite{Doi1988,Nyrkova2007}, which is derived from the
Smoluchowski equation by projecting out longitudinal degrees of
freedom from the mobility tensor, as proposed also in
\cite{Kroy1997}. This amounts to evaluating eq.~(21) of
\cite{Nyrkova2007} with their scaling function replaced with ${\cal
  H}(x)=(1+\dd^2/\dd x^2) \sin(x)/x$. 
An advantage of our approach
over the calculation via the Smoluchowski equation lies in the fact
that the Rotne-Prager corrections can easily be implemented in the 
mobility function. 
Table \ref{cpartab} compares the different
predictions for the constant $C$ in eq.~\eqref{inidecshort}.

\subsection{Logarithmic tails ($t \gg \tau_{e}$)}
\label{glassysf}
Since the dynamic structure factor at long delay times is of the
(incoherent) form
\begin{equation}
  S(q,t) \sim S(q,0) \exp[-q^2\delta r_\perp^2(t)/4]\;,
\end{equation}
its late time behavior is trivially obtained from the dynamic
MSD. Upon exponentiating, Fig.~\ref{fig:msd_gwlc_tube} is translated into
Fig.~\ref{fig:sqt_coll}. Interestingly, the somewhat limited logarithmic
intermediate asymptotics visible in the MSD in the strongly sticky
limit $\EE\gg1$ is thereby extended to much longer times, so that a
logarithmic intermediate decay seems to constitute a quite robust
feature of the dynamic structure factor of semidilute solutions of
sticky semiflexible polymers (cf. Fig. \ref{sqtfit}). The remainder of this paragraph is
therefore dedicated to a closer examination of the analytical
properties of this intermediate asymptotics. For simplicity, we set $\FF=0$ for
the following discussion (the case $\FF >0$ is discussed in appendix
\ref{freevolume}).

We start from the dynamic MSD of a WLC and introduce the stretched
relaxation times $\tilde\tau_n$ eq.~\eqref{newtaun}. For convenience,
we quote the general expression for the MSD by converting eq.~\eqref{msdsum}
to an integral, after taking the limit $L, n_\Lambda \rightarrow \infty$
with $\Lambda = \mbox{const.}$,
\begin{multline}
\delta r_\perp^2(t) = \overline{\delta r_\perp^2(s,s,t)} =
\frac{4 L^3}{\pi^4 \ell_p} \int\limits_0^\infty \frac{\dd n}{n^4}\,
[1-\exp(-t/\tilde\tau_n)].
\label{msdint}
\end{multline}
We perform a change of variables and the dynamic MSD is written as
\begin{equation}
\delta r_\perp^2(t) = \delta r_{\perp,\Lambda}^2(t) + \delta r_{\perp}^{2,G}(t),
\label{rperpgwlc}
\end{equation}
where
\begin{equation}
\delta r_{\perp,\Lambda}^2(t) = \frac{4 \Lambda^3}{\ell_p \pi^4}
\int\limits_1^\infty\frac{\dd n}{n^4}\, \{1-\exp[-(t/\tau_\Lambda) n^4]\},
\label{rperpwlc}
\end{equation}
with $\tau_\Lambda = (\zeta_\perp/\kappa) (\Lambda^4/\pi^4)$ (note that we
assume the simplified treatment of hydrodynamic interactions here, for an improved
discussion of eq.~\eqref{rperpwlc} see appendix \ref{hiapprox}), and
\begin{multline}
\delta r_\perp^{2,G}(t) = \frac{4 \Lambda^3}{\ell_p \pi^4}\\
\times\int\limits_0^1 \frac{\dd n}{n^4}\, \{1-\exp\{-(t/\tau_\Lambda) n^4
\exp[-\EE(1/n-1)]\}\}.
\label{glassyrperp}
\end{multline}
For the slow modes of eq.~\eqref{glassyrperp}, substituting $\overline \ell_\perp(t)
=\Lambda/\pi$ in the friction constant $\zeta_\perp(t)$ is
sufficient.
The first integral can be expressed in terms of an incomplete Gamma function,
\begin{equation}
\delta r_{\perp,\Lambda}^2(t) = \frac{4 \Lambda^3}{3 \ell_p \pi^4}
\left\{1-b (t/\tau_\Lambda)^b \Gamma[-b,(t/\tau_\Lambda)]\right\},
\label{rperplambda}
\end{equation}
where $b=3/4$. (This term is of the same form as the right term in eq.~\eqref{effmedium}.)
The integral saturates in the
limit of long times,
\begin{equation}
\delta r_{\perp,\Lambda}^2
(\infty) = 4 \Lambda^3/3\ell_p \pi^4.
\label{rperpsat}
\end{equation}
The second integral is not immediately solvable analytically. We approximate it within a
certain range of parameters.  Writing
\begin{equation}
\tilde n = \tilde t^{1/4} \exp[-(\EE/4) (1/n-1)] n,
\end{equation}
with the rescaled time $\tilde t=t/\tau_\Lambda$, we approximatively solve the
implicit equation for $n$ and get
\begin{align}
\label{nsub}
n&\approx\frac{1}{1-(4/\EE) \log(\tilde n/n_0 \tilde t^{1/4})}\\
\label{dnsub}
\dd n &\approx\frac{4}{\EE} \frac{\dd \tilde n}{[1-(4/\EE) \log(\tilde n/n_0 \tilde t^{1/4})]^2
\tilde n}.
\end{align}
It is possible to numerically determine the fixed mode number $n_0$. This substitution
is valid as long as the logarithmic term in the denominator of eqs.~\eqref{nsub},
\eqref{dnsub} is not dominant, i.e.\  for $\EE \gg 1$. The remaining integral then reads:
\begin{equation}
\begin{split}
\delta r_\perp^{2,G}(t)& \approx \frac{4 \Lambda^3}{\ell_p \pi^4} \frac{4}{\EE}
\int\limits_0^{\tilde t^{1/4}} \frac{\dd \tilde n}{\tilde n}\, [1-\exp(-\tilde n^4)]\\
&\quad\qquad\qquad\qquad\times\left[1-\frac{4}{\EE} \log(\tilde n/n_0 \tilde t^{1/4})\right]^2\\
&\approx \frac{4 \Lambda^3}{\ell_p \pi^4} (A - B).
\end{split}
\end{equation}
These two integrals can be performed,
\begin{equation}
A=\frac{4}{\EE} \int\limits_0^{\tilde t^{1/4}}
\frac{\dd\tilde n}{\tilde n}\, [1-\exp(-\tilde n^4)]\\
= \frac{1}{\EE} [\gamma_E - \mbox{Ei}(-\tilde t)+ \log(\tilde t)]
\end{equation}
and
\begin{equation}
\begin{split}
B& = \frac{32}{\EE^2} \int\limits_0^{\tilde t^{1/4}} \frac{\dd \tilde n}{\tilde n}\,
[1-\exp(-\tilde n^4)] \log(\tilde n/n_0 \tilde t^{1/4})\\
&\approx \frac{8}{\EE^2} [-1/4 - \log(n_0)] \tilde t + {\cal O}(\tilde t)^2
\qquad (\tilde t \ll 1).
\end{split}
\end{equation}
In the limit $\tilde t/\EE \ll 1$ we can neglect the contribution of
$B$ (which for long times is ultimately approximated by a logarithmic
term in $t$ resulting in corresponding power-law-decay in the dynamic structure factor)
against $A$ and the final result, valid for short and intermediate
times, is provided by eq.~\eqref{gwlclongtime}.
Combining eq. \eqref{sqtlongtime}, and
\eqref{gwlclongtime}, observing that $\mbox{Ei}(-\tilde t)\rightarrow
0\, (\tilde t\rightarrow \infty)$, we arrive at
eq.~\eqref{sfglassylongtime2}.
A comparison of the result eq.~\eqref{gwlclongtime} to the numerically
evaluated expression eq.~\eqref{glassyrperp} is shown
in Figs.~\ref{fig:msd_gwlc_tube}, \ref{fig:sqt_coll}.

As the direct numerical evaluation in Figs.~\ref{sqtfit},
\ref{fig:sqt_coll} shows,
the logarithmic tail of the structure factor for large $\EE$ extends
well beyond the indicated time regime $t \ll \EE \tau_\Lambda$. It
follows from eq.~\eqref{sfglassylongtime2} that the prefactor of the
logarithm (the slope of the logarithmic tail in a semilogarithmic
plot) is inversely proportional to the energy barrier height $\EE$,
which is therefore directly monitored by the slope of the tail of the
structure factor plotted against $\ln t$.  Both $\Lambda$ and $\EE$
can be accurately determined from a fit of eq.~\eqref{sqtmsd} to
experimental data with the $\delta r_\perp^2(t)$ given by the full
model eqs.~\eqref{rperpgwlc}, \eqref{rperplambda},
\eqref{gwlclongtime}.

\section{Conclusions and outlook}
We have presented a thorough theoretical discussion of the dynamic
structure factor of a stiff polymer in a (sticky) solution. We claim
that our predictions, though in qualitative agreement with previous
results, are quantitatively superior and provide the missing link to a
reliable quantitative analysis of microscopic mechanical parameters of
polymers (backbone diameter and persistence length) via dynamic
scattering measurements. Even more important seem the prospects of
applying quasi-elastic scattering as a matchless non-invasive
microrheological technique to explore with high accuracy the so far
poorly understood parameter dependencies of the GWLC stretching
parameter $\EE$. This represents in our opinion one of the most
promising pathways towards a microscopic modeling of the mechanical
properties of biopolymer networks and living cells. A particularly
rewarding application might be the search for an apparently scale
dependent persistence length of biopolymers \cite{Heussinger2007a},
which might arise from the fact that these polymers have a fibrous
substructure requiring a GWLC (rather than a WLC) description already on the 
single polymer level.

\begin{acknowledgement}
We thank R. Merkel for providing us with the experimental light scattering data in
Fig. \ref{sqtfit} and S. Sturm and L. Wolff for a critical reading of the manuscript.
Financial support from the Deutsche Forschungsgemeinschaft (DFG) through FOR 877
(KR 1138/21-1) and from the Leipzig School of Natural Sciences - Building with Molecules
and Nano-objects is gratefully acknowledged.
\end{acknowledgement}
\begin{appendices}

\section{Analytical approximation to hydrodynamic interactions}
\label{hiapprox}
In this section we will derive an analytical approximation for the mode integral
of the MSD of the free GWLC modes, eq.~\eqref{rperpwlc}, which is also the part of
the MSD which corresponds to a static tube. We rewrite
the integral with the help of eq.~\eqref{zetaperpk}:
\begin{equation}
\delta r_{\perp,\Lambda}^2(t)=\frac{4 \Lambda^3}{\ell_p \pi^4} \int\limits_1^{n_0}
\frac{\dd n}{n^4}\, \left\{1-\exp[-(t/\overline\tau_\Lambda) n^4 \log(n_0/n)]\right\},
\label{msdlog}
\end{equation}
where $n_0 \equiv \Lambda \exp(C')/\pi d$ is an upper mode cut-off corresponding
to the finite backbone diameter and $\overline \tau_\Lambda\equiv
(4\pi\eta/\kappa) (\Lambda/\pi)^4$ is the approximate relaxation time of a mode of
wavelength $\Lambda$.
We begin with a substitution of variables, $z^4 \equiv n^4 \log(n_0/n)$. The lower
bound of the integral is then $z_0=\log^{1/4} n_0$. With
$\alpha \equiv 4 z^4/n^4$, $n(z)$ is determined if $\alpha$ is a solution to the equation:
\begin{equation}
\alpha \exp(-\alpha) = 4 \frac{z^4}{n_0^4}.
\end{equation}
Two solutions for $-\alpha$ exist: the two real branches of the Lambert $W$-function. In
our case, the dominant contribution to the integral comes from the mode numbers for which
$n \ll n_0$, this corresponds to $z > n$ or $\alpha \gg 1$. The
relevant solution is therefore given by $\alpha=-W_{-1}(-4 z^4/n_0^4)$. This defines
the upper bound of the integral as $z_1=(4 e)^{-1/4} n_0$. We then have
$W_{-1}(-4 z^4/n_0^4) = -4 z^4/n^4$, and, using the derivative of the Lambert $W$ function,
$W_{-1}'(x)=W_{-1}(x)/\{x [1+W_{-1}(x)]\}$, 
\begin{equation}
\dd n=\sqrt[4]{\frac{4}{\alpha}} \left(1-\frac{1}{1-\alpha}\right) \dd z.
\end{equation}
The Lambert $W$ function may be asymptotically approximated by $W_{-1}(-x)=\gamma + \log(x)$ for $x \ll 1$ with
$|\gamma| \ll |\log(x)|$ \cite{Corless1996}. Here we choose
$\gamma=W_{-1}(-x_0) - \log(x_0)$, i.e., we expand $W_{-1}(-x)-\log(x)$ to $0th$ order
around $x_0 \equiv -4 z_0^4/n_0^4$ (This means, that for very high mode
numbers or very short times the approximation breaks down.)
For the purpose of numerical evaluation, analytical
approximations to $W_{-1}$ may be used \cite{Barry2000}.
Hence we have the following expression for the integral eq.~\eqref{msdlog}:
\begin{multline}
\delta r_{\perp,\Lambda}^2(t)\approx \frac{4 \Lambda^3}{\ell_p \pi^4} 4^{-3/4}
\int\limits_{z_0}^{z_1} \dd z\, \alpha^{3/4}(z) \left(1-\frac{1}{1-\alpha}\right)\\
\frac{1-\exp[-(t/\overline\tau_\Lambda)]}{z^4}.
\label{msdlog2}
\end{multline}
Consider the first factor in the integrand. It is
\begin{equation}
\begin{split}
\alpha^{3/4}(z)&=\left[-W_{-1}\left(-\frac{4 z^4}{n_0^4}\right)\right]^{3/4}\\
&\approx \left[-\gamma(z_0)-\log\left(\frac{4 z^4}{n_0^4}\right)\right]^{3/4}
\quad (z \ll n_0)\\
&\approx\left[-\log\left(\frac{4 z^4}{n_0^4}\right)\right]^{3/4}\\
&\qquad\qquad\qquad\left[1+\frac{3}{4}
\frac{\gamma(z_0)}{\log(4 z^4/n_0^4)}\right].
\label{intfac1}
\end{split}
\end{equation}
We approximate the first factor in eq.~\eqref{intfac1} for $z \ll n_0$ as 
\begin{equation}
\left[-\log(4 z^4/n_0^4)\right]^{3/4}\approx 4^{3/4} \log^{3/4} n_0
\left[1-\frac{3}{4}\frac{\log(\sqrt{2}z)}{\log n_0}\right],
\end{equation}
and the $1/\log(\dots)$ term in the second factor as
\begin{equation}
\begin{split}
\frac{1}{\log(4 z^4/n_0^4)}& \approx \frac{-1}{4 \log n_0
\left[1-\frac{\log(\sqrt{2}z)}{\log n_0}\right]}\\
&\approx \frac{-1}{4 \log(n_0)} \left[1+\frac{\log(\sqrt{2} z)}{\log n_0}\right].
\end{split}
\end{equation}
Taken together, one finds, after grouping the terms according to orders of $\log(z)$:
\begin{multline}
\alpha^{3/4}(z)=4^{3/4} \log^{3/4} n_0 \\
\left[ c_0 + c_1 \log z +
{\mathcal O}(\log z/\log n_0)^2 \right],
\label{intfac1b}
\end{multline}
with
\begin{equation}
\begin{split}
c_0&=1-\frac34 \Bigg\{ \frac{\gamma(z_0)}{4 \log n_0} \left[1+ \frac{\log 2}{8 \log n_0}
\left(1-\frac{3 \log 2}{2 \log n_0}\right)\right] \\
&\qquad\qquad\qquad\qquad\qquad\qquad\qquad\quad+ \frac{\log z}{2 \log n_0}\Bigg\},\\
c_1&=\frac{-3}{4 \log n_0} \left[1+\frac{\gamma(z_0)}{16 \log n_0}
\left(1- \frac{3 \log 2}{\log n_0}\right)\right].
\end{split}
\end{equation}
By arguments similar to above, the second factor in the integrand of
eq.~\eqref{msdlog2} is approximated
for $\alpha \gg 1$ and $z \ll n_0$ as
\begin{equation}
\left(1-\frac{1}{1-\alpha}\right) \approx
1 + \frac{1}{4 \log n_0} + {\mathcal O}(1/\log n_0)^2.
\label{intfac2}
\end{equation}

Consider now the integral
\begin{equation}
\int\limits_{z_0}^\infty \dd z\, \frac{1-\exp[-(t/\overline\tau_\Lambda) z^4]}{z^4} \log z.
\label{intlog}
\end{equation}
With
\begin{equation}
\int\limits_{z_0}^\infty \dd z\,\frac{\log z}{z^4} = \frac{1+3 \log z_0}{9 z_0^3}
\end{equation}
we can (partially) rewrite the integral eq.~\eqref{intlog} in terms of an incomplete Gamma
function (after a change of variables):
\begin{multline}
\frac{1+3 \log z_0}{9 z_0^3} - \frac{1}{16} \Bigg\{\left(
\frac{t}{\overline\tau_\Lambda}\right)^{3/4} \int\limits_{(t/\overline\tau_\Lambda)
z_0^4}^\infty
\dd y\,
\frac{ \exp(-y)}{y^{7/4}} \log y\\
- \log\left(\frac{t}{\overline\tau_\Lambda}\right) \Gamma[-3/4,(t/\overline\tau_\Lambda)
z_0^4]
\Bigg\}.
\label{intlog2}
\end{multline}
The second term can now be integrated by substituting a limit representation
for the logarithm:
\begin{equation}
\begin{split}
\int\limits_a^\infty\dd y\, \frac{\exp(-y)}{y^{7/4}} \log y &= \lim_{n\rightarrow 0}
\int\limits_a^\infty\dd y\, \frac{\exp(-y)}{y^{7/4}} \frac{y^n-1}{n}\\
&=\lim_{n\rightarrow 0}\frac1n [\Gamma(-3/4+n,a)-\\
&\qquad\qquad\qquad\qquad\Gamma(-3/4,a)].
\end{split}
\label{intlog3}
\end{equation}
While the limit exists in terms of a Meijer G function, it can also be performed
numerically by inserting very small values of $n$, which will be sufficient for our
case. With the help of eqs.~\eqref{rperpwlc}, \eqref{intfac1b}, \eqref{intfac2},
\eqref{intlog2}, ~\eqref{intlog3}, eq.~\eqref{msdlog2} can now be
fully approximated and the final result valid for times $t \gg (4 \pi \eta/\kappa) d^4$
is given in eq. ~\eqref{msdlogtot}. A comparison of this result with the simple
approximation given in the main text is shown in Fig.~\ref{msdcomp}.

\begin{strip}

\vspace{1cm}
\noindent\parbox{.5\textwidth}{\hfill\vrule height4pt\hrule}
\begin{multline}
\delta r_{\perp,\Lambda}^2(t) = \frac{4 \Lambda^3}{\ell_p \pi^4} \log^\frac34 n_0
\left(1+\frac{1}{4 \log n_0}\right)
\Bigg\{ \frac{c_0}{3 z_0^3} \left\{ 1- \frac{3 z_0^3}{4}
\left(\frac{t}{\overline\tau_\Lambda}\right)^\frac34 \Gamma[-3/4,(t/\overline\tau_\Lambda)
z_0^4]
\right\}\\
+ c_1 \Bigg\{ \frac{1+3 \log z_0}{9 z_0^3}-\frac{1}{16}
\left(\frac{t}{\overline\tau_\Lambda}\right)^{\frac34}
\left\{\lim_{n \rightarrow 0} \frac{1}{n} \{
\Gamma[-3/4+n,(t/\overline\tau_\Lambda) z_0^4]-\Gamma[-3/4,(t/\overline\tau_\Lambda)
z_0^4] \}
 - \log\left(\frac{t}{\overline\tau_\Lambda}\right) \Gamma[-3/4,(t/\overline\tau_\Lambda)
z_0^4] \right\}
\Bigg\} \Bigg\}.
\label{msdlogtot}
\end{multline}
\hfill\parbox{.5\textwidth}{\hrule\vrule height2pt depth2pt}%
\end{strip}

\section{Approximation to free volume contributions}
\label{freevolume}
In this section we calculate the MSD of a GWLC with purely steric interactions, in the
special case $\FF>0$, $\EE=0$. We may start from eq.~\eqref{msdint}, where we introduce
the stretched relaxation times $\tilde\tau_n$ for the free volume contributions
according to eq.~\eqref{taunfreevolume}. The integral is split into the contributions
due to the free high modes $\delta r_{\perp,L_e}^2(t)$ and due to the slow modes
$\delta r_{\perp}^{2,G}(t)$, as in eq.~\eqref{rperpgwlc}. We have
\begin{multline}
\delta r_\perp^{2,G}(t) =\\
\frac{4 L_e^3}{\ell_p \pi^4} \int\limits_0^1
\frac{\dd n}{n^4} \left\{ 1 - \exp\{-(t/\tau_e) n^4 \exp[-\FF (1/n^4 -1)]\}\right\},
\label{rperpglassyfreevolume}
\end{multline}
where we substituted $\Lambda \rightarrow L_e$. We will
approximate this integral, by
a change of variables,
$\exp(z) =\tilde t n^4 \exp[-\FF$ $ (1/n^4 -1)]$, with the rescaled time
$\tilde t\equiv t/\tau_e$, which implies
\begin{equation}
n^4= \FF/\,W\left[\tilde t \FF \exp(\FF-z)\right],
\end{equation}
with $W$ being the Lambert function. Using the approximation $W(x)\approx \log(1+x)$
\cite{Barry2000},
which becomes exact in the limit $x \rightarrow 0$,
\begin{multline}
\dd n \approx\frac{\FF^{1/4}}{4 \log^{5/4}\left[1+\tilde t \FF \exp(\FF-z)\right]}\\
\times\left[1-\frac{1}{1+\tilde t \FF \exp(\FF-z)}\right] \dd z.
\end{multline}
Now eq.~\eqref{rperpglassyfreevolume} may be rewritten,
\begin{equation}
\begin{split}
\delta r_\perp^{2,G}&\approx\frac{4 L_e^3}{\ell_p \pi^4} \frac{1}{\FF^{3/4}}\\
&\times \frac{1}{4}\left[\int\limits_{-\infty}^{0} \dd z\, f(\tilde t \FF e^\FF,z)+
\int\limits_0^{\log \tilde t} \dd z\, f(\tilde t \FF e^\FF,z)\right]\\
&\equiv A + B,
\label{freevolumesubst}
\end{split}
\end{equation}
with
\begin{equation}
f(x,z)\equiv \frac{1-\exp[-\exp(z)]}{\log^{1/4}[1+x/\exp(z)]}
\times\left[1-\frac{1}{1+x/\exp(z)}\right].
\label{freevolumeint}
\end{equation}
We will in the following
restrict the discussion to the case $\FF \lesssim 1$ and $\tilde t \gg \exp(-\FF)/\FF$. In this case
we may approximate the first term in eq.~\eqref{freevolumesubst} by the
following integral:
\begin{multline}
A\approx\frac{4 L_e^3}{\ell_p \pi^4} \frac{1}{\FF^{3/4}}
\frac{1}{4}\int\limits_{-\infty}^0 \dd z\,
\frac{\exp(z)}{\log^{1/4}[\tilde t \FF \exp(\FF-z)]}\\
\times\left[1-\frac{1}{\tilde t \FF \exp(\FF-z)}\right].
\end{multline}
Here we expanded the double exponential factor to first order in $\exp(z)$. The integral is done
analytically and the asymptotic result is
\begin{equation}
A \propto \frac{1}{4 \log^{1/4}[ \tilde t \FF \exp(\FF)]}\qquad (\tilde t \FF \exp(\FF) \rightarrow \infty)
\end{equation}
We observe that the double exponential factor in eq.~\eqref{freevolumeint} may be approximated
by a step function, $\theta(z)$.  The remaining integral of eq.~\eqref{freevolumesubst}
can then be performed analytically, and we arrive at eq.~\eqref{msdglassyfreevolume} of the main text.
To assess the validity of the approximation made, we first remark that we require $\log \tilde t \gg 1$.
We then compare the two terms of 
eq.~\eqref{freevolumesubst} in the limit $\tilde t \FF \exp(\FF) \rightarrow \infty$ and find
\begin{equation}
A/B \sim \frac{3}{4} \log^{-1}[\tilde t \FF \exp(\FF)] \qquad (\tilde t \FF \exp(\FF) \rightarrow \infty),
\end{equation}
i.e., the approximation leading to eq.~\eqref{msdglassyfreevolume}
is valid for $\log[\tilde t \FF \exp(\FF)] \gg 1$.

A comparison of this result to the numerically evaluated
expression eq.~\eqref{rperpglassyfreevolume} is shown
in Figs.~\ref{fig:msd_gwlc_tube}, \ref{fig:sqt_coll}.
Though eq.~\eqref{msdglassyfreevolume} is only valid
asymptotically, qualitative agreement is already reached for $\tilde t \gg \exp(-\FF)/\FF$, $\FF \lesssim 1$.

\end{appendices}


\end{document}